\documentclass[preprint]{aastex}

\newcommand{\HI}{\mbox {\sc H\thinspace{i}}}
\newcommand{\kms}{\mbox{km\, s$^{-1}$}}
\newcommand{\cmsq}{\mbox{cm$^{-2}$}}
\newcommand{\Jykms}{\mbox{Jy\, km\, s$^{-1}$}}
\newcommand{\iraf}{\mbox{{\sc {iraf}}}}
\newcommand{\hipass}{\mbox{HIPASS}}
\newcommand{\miriad}{\mbox{\sc miriad}}

\newcommand{\kkms}{\mbox{K\,km\, s$^{-1}$}}
\newcommand{\kpc}{\mbox{kpc}}
\newcommand{\acmsq}{\mbox{atoms\, cm$^{-2}$}}
\newcommand{\sqcm}{\mbox{cm$^{-2}$}}
\newcommand{\nhi}{\mbox{$N_{\rm HI}$}}

\newcommand{\Mpc}{\mbox{Mpc}}
\newcommand{\mJy}{\mbox{mJy}}
\newcommand{\Msun}{\mbox{${\cal M}_\odot$}}

\newcommand{\erg}{\mbox{ergs s$^{-1}$ cm$^{-2}$ sr$^{-1}$ Hz$^{-1}$}}

\slugcomment{Version 9-May-2000}

\shorttitle{An Extragalactic HI Cloud}
\shortauthors{Kilborn et al.}

\begin{document}


\title{An Extragalactic \HI\ Cloud with No Optical Counterpart?}


\author{
V. A. Kilborn,\altaffilmark{1,13}
L. Staveley-Smith,\altaffilmark{2}
M. Marquarding,\altaffilmark{1} 
R. L. Webster,\altaffilmark{1}
D. F. Malin,\altaffilmark{3}
G. D.  Banks,\altaffilmark{4}   
R. Bhathal,\altaffilmark{5} 
W. J. G. de Blok,\altaffilmark{2} 
P. J. Boyce,\altaffilmark{4} 
M. J. Disney,\altaffilmark{4} 
M. J. Drinkwater,\altaffilmark{1}
R. D. Ekers,\altaffilmark{2} 
K. C. Freeman,\altaffilmark{6} 
B. K. Gibson,\altaffilmark{7} 
P. A.  Henning,\altaffilmark{2,8} 
H. Jerjen,\altaffilmark{6}
P. M. Knezek,\altaffilmark{9,14}
B. Koribalski,\altaffilmark{2} 
R. F. Minchin,\altaffilmark{4,14}  
J. R. Mould,\altaffilmark{6} 
T. Oosterloo,\altaffilmark{10} 
R. M. Price,\altaffilmark{2,8} 
M. E. Putman,\altaffilmark{6} 
S. D. Ryder,\altaffilmark{11} 
E. M. Sadler,\altaffilmark{12}  
I. Stewart,\altaffilmark{2} 
F. Stootman,\altaffilmark{5} 
and A. E. Wright\altaffilmark{2}
}

\altaffiltext{1}{University of Melbourne, School of Physics, Parkville, 
Victoria 3052, Australia}

\altaffiltext{2}{Australia Telescope National Facility, CSIRO, P.O. Box 76,
Epping, NSW 2121, Australia}

\altaffiltext{3}{Anglo-Australian Observatory, P.O. Box 296, 
Epping, NSW 2121,  Australia}

\altaffiltext{4}{University of Wales, Cardiff, Department of Physics 
\& Astronomy, P.O. Box 913, Cardiff CF2 3YB, U.K.}

\altaffiltext{5}{University of Western Sydney Macarthur, Department of 
Physics, P.O. Box 555, Campbelltown, NSW 2560, Australia}

\altaffiltext{6}{Research School of Astronomy and Astrophysics, 
ANU, Weston Creek P.O., Weston, ACT 2611, Australia.}

\altaffiltext{7}{Center for Astrophysics and Space Astronomy, University of 
Colorado, Boulder CO, 80309-0389, USA.}

\altaffiltext{8}{University of New Mexico, Department of Physics \& 
Astronomy, 800 Yale Blvd. NE, Albuquerque, NM 87131, USA}

\altaffiltext{9}{Space Telescope Science Institute, 3700 San Martin
Drive, Baltimore, MD, 21218, USA}

\altaffiltext{10}{Istituto di Fisica Cosmica, via Bassini 15, I-20133, 
Milano, Italy }

\altaffiltext{11}{Joint Astronomy Center, 660 North Aohoku Place, Hilo, 
H{\sc i} 96720, USA}

\altaffiltext{12}{University of Sydney, Astrophysics Department, School 
of Physics, A28, Sydney, NSW 2006, Australia}

\altaffiltext{13}{vkilborn@physics.unimelb.edu.au}

\altaffiltext{14}{Visiting Astronomer, Cerro Tololo Inter-American
Observatory, National Optical Astronomy Observatories, which is
operated by the Association of Universities for Research in Astronomy,
Inc. (AURA) under cooperative agreement with the National Science Foundation.}



\begin{abstract}

  We report the discovery, from the \HI\ Parkes All-Sky Survey
  (HIPASS), of an isolated cloud of neutral hydrogen which we believe
  to be extragalactic. The \HI\ mass of the cloud (HIPASS J1712-64) is
  very low, $1.7 \times 10^7 \Msun$, using an estimated distance of
  $\sim 3.2$ Mpc. Most significantly, we have found no optical
  companion to this object to very faint limits ($\mu(B)\sim 27$ mag
  arcsec$^{-2}$). HIPASS J1712-64 appears to be a binary system
  similar to, but much less massive than, HI 1225+01 (the Virgo \HI\ 
  Cloud) and has a size of at least 15 kpc. The mean velocity
  dispersion, measured with the Australia Telescope Compact Array
  (ATCA), is only 4 \kms\ for the main component and because of the
  weak or non-existent star-formation, possibly reflects the thermal
  linewidth ($T<2000$ K) rather than bulk motion or turbulence. The
  peak column density for HIPASS J1712-64, from the combined Parkes
  and ATCA data, is only $3.5\times 10^{19}$ \cmsq, which is estimated
  to be a factor of two below the critical threshold for star
  formation. Apart from its significantly higher velocity, the
  properties of HIPASS J1712-64 are similar to the recently recognised
  class of Compact High Velocity Clouds. We therefore consider the
  evidence for a Local Group or Galactic origin, although a more
  plausible alternative is that HIPASS J1712-64 was ejected from the
  interacting Magellanic Cloud/Galaxy system at perigalacticon $\sim
  2\times 10^8$ yr ago.

\end{abstract}




\section{INTRODUCTION}
\label{sec:intro}

Until recently, there has been no extensive survey of the
extragalactic sky in neutral hydrogen (\HI). The mass distribution in
the Universe has therefore been traced by galaxy surveys in other
wavebands, particularly in the optical, but also in the infrared using
the IRAS satellite. Because of the limited nature of existing \HI\
surveys, it is unclear whether there is a significant local population
of \HI-rich objects not seen in optical and infrared surveys (away
from the Galactic Plane). In particular, the existence of \HI\ clouds
without optical counterparts, or '\HI\ protogalaxies' might significantly
add to the local \HI\ mass density of the Universe. Recent
observations by \citet{zwa97} and \citet{sch98} arrive at somewhat
differing conclusions regarding the low \HI-mass galaxy population,
leading to some uncertainty in the contribution of these objects to
the HI mass in the local Universe.

Limits to the presence of a population of \HI-rich clouds and
`protogalaxies' have been set by a number of `blind' or semi-`blind'
\HI\ surveys conducted in a range of environments: clusters
\citep{bar97, dic97}, groups \citep{kra99, banks99, fis81a}, voids
\citep{kru84}, and in the general field \citep{hen95, sor94, spi98,
  sho77}. In addition, there have been numerous observations in
off-source calibration regions for galaxies which have been
optically-selected.  For example, \citet{fis81b} serendipitously
detected a couple of dozen galaxies in the calibration scans from
their survey, all with optical counterparts.

A number of \HI\ clouds without optical counterparts have been found.
However, many appear to be closely associated with massive galaxies
\citep{wil88, gio95} and perhaps the product of tidal interactions
\citep{hvg96}. A number of potential \HI\ protogalaxies have later
been classified as High Velocity Clouds \citep{mat75}, and argument
continues over the extragalactic nature of a specific class of Compact
High Velocity Clouds \citep{bb99, zwa00}. One of the larger blind \HI\
surveys to date \citep{spi98} detected 75 galaxies in 55 deg$^2$ of
sky. One detection was found not to have any obvious optical
counterpart, though a bright star just near the peak of the \HI\
emission may obscure a low surface brightness optical companion.

The closest and best examples of \HI\ clouds without prominent
counterparts are the Virgo Cloud HI 1225+01 \citep{gio89}, and the Leo
ring \citep{sch83}. Both of these \HI\ clouds were later found to be
associated with optical galaxies.  HI 1225+01 was found
serendipitously during routine calibration observations. It comprises
two regions of \HI\ emission: the largest has a Magellanic-type dwarf
irregular galaxy at the position of the highest \HI\ column density,
while the smaller, possibly infalling cloud shows no optical emission
to a faint limiting magnitude \citep{che95}.  The larger component has
a total dynamical mass of $\sim 1.0 \times 10^{10}\,d_{20}\Msun$, and
a gas mass of $\sim 3 \times 10^9\,d^2_{20}\Msun$ where $d_{20}$ is
the distance in units of $20\,\Mpc$.  The heliocentric velocity of the
cloud is 1275 \kms\, and the extent of the \HI\ emission is about
$\sim 200\,d_{20}\kpc$.  The Leo ring is an intergalactic cloud found
during \HI\ observations of the Leo group of galaxies.  It comprises a
large ring of \HI\ emission surrounding NGC 3384 and M105, with an
\HI\ mass of $\sim 10^9 \Msun$.  The velocity structure suggests that
it is probably not a tidal tail from the nearby galaxies, but a
primordial gas cloud which has not started forming stars
\citep{sch89}. The cloud is gravitationally bound to the optical
galaxies in the group. Both of these objects have been used in
H$\alpha$ studies to determine limits on the local ionising background
\citep{don95}, finding limits which suggest that quasar light, not
galactic light, dominates the local ionizing background at low
redshift.

The \HI\ Parkes\footnote{The Parkes telescope is part of the Australia
  Telescope which is funded by the Commonwealth of Australia for
  operation as a National Facility managed by CSIRO. } All Sky Survey
(\hipass) has been underway since 1997 and, when complete, will be the
largest blind \HI\ survey to date, surveying a volume at least two
orders of magnitude larger than any previous survey. In early \hipass\ 
observations of 600 deg$^2$ in the Centaurus region,
\citet{banks99} found 10 new members of the nearby Centaurus A group.
However, all of these have optical companions, although 5 are faint and
were previously uncatalogued. Other early observations include the south
celestial cap \citep{put98, kws99}. During routine inspection of this
data, a resolved \HI\ detection, \hipass\ J1712-64, was selected for
follow-up observations as it showed no catalogued galaxy in the vicinity
and no optical counterpart on the Digitised Sky Survey. It
also had a low systemic velocity (though well-separated from Galactic
gas and High Velocity Clouds) and a significant angular size. 

The observations of \hipass\ J1712-64 are discussed in
\S~\ref{sec:obs}; the \HI\ structure, dynamics and optical limits are
discussed in \S~\ref{sec:phys}. Alternative High Velocity Cloud and
Local Group hypotheses are discussed in \S~\ref{sec:alt}, the nature
of HIPASS J1712-64 is discussed in \S~\ref{sec:dis}, and the results are
summarised in \S~\ref{sec:summ}.  Throughout we assume a Hubble
constant of 75 \kms Mpc$^{-1}$.

\section{OBSERVATIONS AND DATA REDUCTION}
\label{sec:obs}

\subsection{\hipass}
\label{sec:hipass}

\hipass\ is a blind \HI\ survey of the sky $\delta <2\arcdeg$. The
survey uses the Parkes 64 m telescope which is equipped with a 13-beam
receiver \citep{sta96}. The observations began in February 1997, and
will be finished by the start of 2000. A northern extension is being
contemplated from Parkes, and parts of a complementary northern survey
are currently being conducted at Jodrell Bank Observatory's Lovell
Telescope. The velocity range covered is $-1200$ to $12700~\kms$.
Observations are taken by scanning the multibeam receiver in
declination strips of length $\sim$ 8\arcdeg. Each declination scan is
separated by 35\arcmin\ in R.A., and each area of sky is
scanned five times, resulting in a final scan separation of 7\arcmin.
The final integration time is approximately 460 s beam$^{-1}$. The
mean telescope FWHM beam is 14\farcm3, although the gridding process
increases this to $\sim 15\farcm5$ \citep{bar00}.

Bandpass calibration, spectral smoothing and Doppler correction of the
data are applied in real time at the telescope \citep{bar98}.  Once
all data are collected, the spectra are gridded into data cubes which
have an rms noise level of $\sim 13 \, \mJy$ beam$^{-1}$. The channel
spacing in the final cubes is $13.2 \ \kms$, the FWHM resolution is
$18.0 \ \kms$ and the pixel size is $4 \arcmin \times 4 \arcmin$.

\subsection{\HI\ Observations}

\hipass\ J1712-64 was discovered during a routine visual inspection of
data cubes. The detection is very significant: it has a peak flux of
$\sim 150\, \mJy$ beam$^{-1}$, corresponding to a signal-to-noise
ratio of about 12. It is also significantly resolved in the beam of
the Parkes telescope, extending over at least two beamwidths.  On 1999
June 2, a further 2.5 hour observation was made at the Parkes
telescope, using the multibeam receiver, to confirm the detection. The
observation was of a $3\arcdeg \times 4\arcdeg$ field centered on the
original detection. This observation also resulted in a clear positive
detection. The rms noise in the follow-up observation is $19\, \mJy$
beam$^{-1}$.  The combined, and spatially integrated, Parkes spectrum
is shown in Fig.~\ref{f:spec}.  The total flux density peaks at 250
mJy at a heliocentric velocity of $\sim$ 455 \kms.

Higher resolution observations in the \HI\ line were made at the
Australia Telescope Compact Array (ATCA) on 1999 June
30. Approximately 12 hours of data was obtained using the 375~m array,
and the pointing center for the observations was R.A. $17^{\rm
h}12^{\rm m}13^{\rm s}$, Decl. $-64^{\rm d}39^{\rm m}12^{\rm s}$
(J2000).  The secondary calibrator PKS 1814-637 was observed every 30
minutes for 4 minutes, to calibrate the amplitude and phase of the
visibility data. The primary flux scale was set using the primary
calibrator PKS 1934-638. The bandwidth of 8 MHz was divided into 512
channels, resulting in a channel spacing of 3.3 \kms\ and a FWHM
resolution, before any smoothing, of 4.0 \kms.

The data were reduced using the \miriad\ reduction package. The data
were edited, calibrated in amplitude and phase and bandpass-corrected.
Several bright continuum sources were then removed by fitting spectral
baselines to the line-free channels in the visibility domain. Natural
weighting was used to form the data cube, which was {\sc clean}ed
until the absolute maximum residual in each plane fell below 10 mJy.
The final FWHM beam is $2\farcm0 \times 1\farcm9$, and the rms noise
is 3.7 mJy beam$^{-1}$. No correction for the primary beam pattern was
applied.

J1712-64 was again detected in the ATCA data. The spatially integrated
spectrum (smoothed to the same velocity resolution) is overlaid on the
Parkes/\hipass\ data in Fig.~\ref{f:spec}.  All except $\sim20$ \% of
the Parkes flux density (mainly at higher velocities) is recovered.
This implies that some diffuse or low-level emission is being missed
by the ATCA. The ATCA channel images (Hanning-smoothed to 6.6 \kms\ 
resolution) are shown in Fig.~\ref{f:chan}.

A combined cube of the Parkes and ATCA data set was made by spectrally
smoothing the ATCA data to match the \hipass\ data, then spatially
resampling the \hipass\ data to match the ATCA data.  The data were
then combined using the \miriad\ task {\sc immerge}.  It was assumed
that both the ATCA and \hipass\ flux density scales were accurate to a
few per cent. The final column density image, with contours overlaid
is shown in Fig.~\ref{f:mom}. The ATCA velocity field, derived from
the {\sc gipsy} task {\sc moments}, is shown in Fig.~\ref{f:mom1}
overlaid on the column density image.

\subsection{Optical Observations}

A search for an optical counterpart using blue and red film copies of
the UKST/ESO southern sky survey revealed no optical
counterpart. Subsequently, two deep original plates were obtained from
the UK Schmidt Telescope (UKST) plate archive and subjected to a
photographic enhancement technique \citep{mal78} which is capable of
revealing extended features over 5 mag fainter than the night sky
\citep{mh97}. At the site of the UKST (Siding Spring) this is
typically $\mu(B) = 22.5$ mag arcsec$^{-2}$.

Only two suitable plates were available, and an image was made by
combining photographically enhanced derivatives from both of them:
J1659 (IIIa-J/GG 395, 395--530 nm) and OR15092 (IIIa-F/RG 590,
590-690 nm). Both were plates of excellent quality but in both cases
the object centre was close to the vignetting region towards the edge
of the UKST focal plane.

The IIIa-J passband is the most sensitive to extended stellar light
(assuming a solar spectrum) and would reveal extended features at
least as faint as 27.5 mag arcsec$^{-2}$. The derivative from the
red-sensitive plate achieves a detection limit of about $\sim$ 26.8
mag arcsec$^{-2}$ because of the brighter night sky in the
red. Nothing unusual is seen on deep derivatives from either plate, or
on the combined image. However, it should be mentioned that large,
featureless low surface brightness objects are particularly difficult
to detect in crowded fields. For an optical companion of 1\arcmin\ in
diameter and lying just beneath our optical limit, these limits suggest an
absolute magnitude $M_B>-8.8$ mag, and an \HI\ mass-to-light ratio
$M_{HI}/L_B>24$ (Table~\ref{table2}) . 

The source was also imaged in the optical $I$-band at the ANU 40-inch
telescope at Siding Springs Observatory on 1999 April 15.  It was also
imaged in the B and R-bands by the CTIO 40-inch telescope on 1999 May 4.
The data were reduced using the \iraf\footnote{\iraf\ is distributed by
the National Optical Astronomy Observatories, which is operated by the
Association of Universities for Research in Astronomy, Inc., under
contract to the National Science Foundation.} package and calibration
for the $B$ and $R$-band images used photometric observations of the
Landolt Standard field 104 \citep{lan83}.  The optical point source
magnitude limits are $B\sim 19.8$ and $R\sim 19.2$ from the CCD
observations.  None of the images show any evidence of extended
optical emission near the location of the \HI\
detection. Fig.~\ref{f:olay} shows the \HI\ contours of \hipass\
J1712-64 overlayed on the combined $B$ and $R$-band CCD images, showing a
dense starfield but no extended objects.

\section{PHYSICAL PARAMETERS}
\label{sec:phys}

\subsection{\HI\ Structure}

HIPASS J1712-64 is well-resolved in each of the velocity channels with
significant emission (see Fig. \ref{f:chan}). The object appears to
have two major components. The prominent one (hereafter the NE
component) is centered at R.A.  $17^{\rm h}12^{\rm m}35^{\rm s}$,
Decl.  $-64^{\circ}38\arcmin12\arcsec$ (J2000) and has a velocity
spread from 438 to 464 \kms\ in the ATCA data.  The systemic
velocity is $451\ \kms$ and the velocity width is $W_{50} = 11\ \kms$
(Table~\ref{table1}). The velocities appear to increase from southeast
to northwest across this component, but the difference is only a few
\kms\ at the most (Fig.~\ref{f:mom1}). The other component (hereafter the SW
component) is centered at R.A.  $17^{\rm h}11^{\rm m}35^{\rm s}$,
Decl.  $-64^{\circ}45\arcmin11\arcsec$ (J2000) and has a velocity
spread from 451 to 477 \kms. The systemic velocity is $464\ 
\kms$ and the velocity width is $W_{50} = 20\ \kms$
(Table~\ref{table1}). Any gradient across this component is again only
a few \kms\ (Fig.~\ref{f:mom1}).

A bridge of emission appears to join the NE and SW components in a
manner reminiscent of the Virgo Cloud HI 1225+01 \citep{che95}.
Velocities appear to increase smoothly from the NE to the SW
components. The overall \HI\ structure appears to be that of two
separate, but interacting components.

\subsection{Optical Limits}
\label{sec:op}

The source is located close to the Galactic plane at $(l, b) =
(326\fdg6, -14\fdg6$), and therefore extinction may be important.  We
have estimated this using three methods.  The local \HI\ column
density from the \hipass\ data (reprocessed to avoid any spatial
filtering) is $484\ \kkms$, which corresponds to $\nhi = 8.8 \times
10^{20}\ \acmsq$.  Following the conversion formula of \citet{bh78},
this implies $E(B-V) =0.12$ mag.  The extinction determined from the
COBE/DIRBE and IRAS/ISSA maps based on dust temperature is essentially
identical: $E(B-V) = 0.11$ mag \citep{schl98}. Finally, the
\citet{bh82} extinction maps suggest a similar value $E(B-V) =0.09$
mag.  Taking the mean value, $E(B-V) =0.11$, this implies a blue-light
absorption of $A_B = 4.0 E(B-V) = 0.44$ mag.  Thus although the
stellar density in the optical images is quite high (see
Fig.~\ref{f:olay}), there is no evidence for large optical extinction
in the direction of the \HI\ detection.

An optical counterpart could be overlooked if the J1712-64 is very
close to the Milky Way and resolved into stars.  One method of
detecting a resolved galaxy in a crowded field of stars is to search a
colour-colour plot of the stars in a field, to look for a separate
stellar population. This method was used to detect the Sagittarius
dwarf galaxy \citep{iba95}. We applied this method to the $B$ and
$R$-band CCD images.  These images cover an area of about 1 deg$^2$,
which is substantially larger than the \HI\
detection. Fig~\ref{f:hist} shows a histogram of stellar magnitudes
for a 20\arcmin\ field centred on the \HI\ detection, and other nearby
fields of the same size. These fields are statistically equivalent,
thus there is no evidence for a resolved galaxy in the direction of
the \HI\ detection.

\subsection{Distance and Mass}

The systemic velocity of the \HI\ detection is $v_{\rm sys} = 451$
\kms\ (NE component). The velocity with respect to the Local Group can
be determined using $v_{\rm LG} = v_{\rm sys} -79 \cos l \cos b +
296\sin l \cos b -36 \sin b$ \citep{yts77}. This gives $v_{\rm LG} =
239$ \kms\ which, if we assume \hipass\ J1712-64 to be extragalactic
and $H_{\circ}=75$ km~s$^{-1}$~Mpc$^{-1}$, corresponds to a
distance of 3.2 Mpc (see Table~\ref{table2}). This would place
\hipass\ J1712-64 at some distance beyond the Local Group. This
distance is subject to considerable uncertainty. Alternate Local Group
velocities range from $v_{\rm LG}=291\ \kms$ (using $300 \sin \ell
\cos b$) to $v_{\rm LG} =172$ \kms\ ({\sc potent}, \citet{ber90}),
giving distances of 3.9 Mpc and 2.3 Mpc, respectively. Without the aid
of optical distance measures, subsequent calculations involving the
distance to J1712-64 will therefore be based on the value of $3.2\ 
(\pm 1)$ Mpc.  At this distance, the projected separation of the NW
and SW components is $\sim 9\ \kpc$, the overall \HI\ size is $\sim
15\ \kpc$ and the \HI\ mass is $\sim1.7 \times 10^7 \Msun$. For a
substantially smaller distance, say 100 kpc (see \S~\ref{sec:alt}), the
separation, size and mass are $\sim 300$ pc, $\sim 500$ pc and
$\sim1.7 \times 10^4 \Msun$, respectively.

The velocities within J1712-64 are very low. The velocity separation
of the NE and SW components is only 12 \kms, and the velocity
dispersions within each component are small. For the NE component, the
mean velocity dispersion from the ATCA data is $\sigma \sim 4\ \kms$\ 
(see Table~\ref{table1}), possibly slightly more for the SW component,
although the S/N ratio is too low for a reliable estimate.

If the velocity difference between the NE and SW components is
indicative of the rotation of a bound, binary system at 3.2 Mpc, the
minimum system mass is $1.5\times 10^8 \Msun$ \citep{fg79}, an order
of magnitude above the \HI\ mass. The `virial' distance, at which the
total mass is 1.3 times the \HI\ mass (to account for He), is
implausibly distant ($> 20$ Mpc) implying that the system contains
substantial dark matter ($\sim 9 M_{\rm HI}$, which would be a 
reasonable value for
disk galaxies), or else is not bound. The orbital period of
the two components corresponding to the minimum total mass is 
$\sim 4\ (d/{\rm 3.2~Mpc})$ Gyr. 

The NE component alone can be modelled as a pressure-supported sphere of gas
of total enclosed mass

\begin{equation}
M_T = \frac {\alpha \sigma^2 r}{G}
\end{equation}
where $\sigma$ is the line-of-sight velocity dispersion ($\sim 4\ 
\kms$), $r$ is the radius ($\sim 4$ kpc) and $\alpha \sim 1-3$
depending on the mass distribution. For an isothermal sphere, $\alpha
\approx 2$, which gives $M_T({\rm NE}) = 3.0 \times 10^7 \Msun$, which
is only $\sim 1.8$ times higher than the \HI\ mass for this component.
Again, for a distance of 100 kpc, the mass drops to $M_T({\rm NE}) =
10^6 \Msun$, but this is now $\sim 60$ times greater than the \HI\ mass if
in hydrostatic equilibrium.

\subsection{Stability}

The critical column density above which local axisymmetric
instabilities can occur in a uniformly rotating gaseous disk is
given by the Toomre stability criterion:

\begin{equation}
\Sigma_c = \frac{2 v_s \Omega}{\pi G},
\end{equation}
where $v_s$ is the sound speed and $\Omega$ is angular frequency
\citep{bt87}. \citet{ken89} has shown that similar relations for
differentially rotating stellar disks approximately predict the column
density above which star formation occurs in spiral galaxies. If we
use $\Omega= v/r$ with $v\approx\sigma\approx v_s$ and $r=4$ kpc for
the NE component of J1712-64 (assuming it to be extragalactic), then
we obtain $\Sigma_c \approx 8\times 10^{19}\ \acmsq$.  This is a
factor of two above the peak measured column density of just $3.5
\times 10^{19}\ \acmsq$ (Fig~\ref{f:mom}), consistent with the absence
of visible star-formation.

\section{ALTERNATIVES}
\label{sec:alt}

\subsection{A Local Group Galaxy?}
\label{sec:lggal}

Nearby, faint galaxies which are resolved into stars can be especially
hard to spot even far from the crowding in the Galactic Plane (e.g.
Sextans -- \citet{irw90}).  In \S~\ref{sec:op} we have shown that
there is no evidence in color-magnitude diagrams for this. In
addition, the Local Group velocity, $v_{\rm LG}=239$\ \kms\ is well beyond
the $\pm 60$\ \kms\ which appears to encompass most known Local Group
members \citep{gre97,ber1999}. This is demonstrated in
Fig.~\ref{f:lg}, where J1712-64 has the highest heliocentric and Local
Group velocity of all the probable and possible Local Group members
plotted. It has also has a higher velocity than the class of `Local Group
outliers' which have distances of $1\sim 2$ Mpc and are regarded by
\citet{gre97} as only potential members. 

In addition, the irregular galaxy IC 4662 lies 3\fdg7 on the sky from
\hipass\ J1712-64, has a velocity of $v_{\rm sys}=302\pm 1\ \kms$,
almost 150 \kms\ lower, yet is not classified as belonging to the
Local Group \citep{ber1999}.  Therefore, from its velocity, there is
no evidence for J1712-64 being a galaxy lying within the Local Group.

\subsection{A High Velocity Cloud?}

The most likely alternative explanation for the classification of
\hipass\ J1712-64 is that it is an unusual high-velocity cloud (HVC)
lying at the edge of our own Galaxy. HVCs are \HI\ objects which do
not fit into a simple model of Galactic rotation and do not have
optical counterparts \citep{ww97,bb99}.  They are widespread across the
sky, with a covering factor estimated to be 0.2 -- 0.4 for $N_{\rm
  HI}>7 \times 10^{17}\ \acmsq$ \citep{ww97}.  There are two good
reason to exclude the possibility of J1712-64 being a normal HVC.  The
first is its velocity, which is 40 \kms\ higher than any currently
catalogued HVC (e.g. those around the Large Magellanic Cloud --
\citet{mbap00}).  The second is J1712-64's HI structure.  Most HVCs
have a core-halo morphology with only 10--50\% of the total flux able
to be recovered by interferometry \citep{ww97}. As shown in
Fig.~\ref{f:spec}, however, the ATCA observations were able to recover
80\% of the total flux. It may, however, be an extreme version of
the compact HVCs studied in detail by \citet{bb00}. These objects
exhibit narrow linewidths and also show apparent velocity gradients.

J1712-64 lies in a 2400 deg$^{2}$ region around the south celestial
cap \citep{put98} where many of the HVCs appear to have originated
from the interaction of the Magellanic Clouds with the Milky Way. It
is possible that a compact HVC similar to J1712-64 could have
originated from gravitational scattering off the three-body Magellanic
Clouds/Galaxy system.  However, \hipass\ J1712-64 does not lie close
to the Clouds (it is 46\arcdeg\ from the LMC), and the Leading Arm,
which includes HVCs 334 and 352 of \citet{ww91} (see Figure~3 of
\citet{put98}), is $\sim 20\arcdeg$ away.  A velocity separation of
$\sim 300\ \kms$ and a spatial separation from the LMC of $\sim 60$
kpc (or $\sim 80$ kpc from the Galaxy) would be consistent with an
ejection event coinciding with the last LMC/SMC encounter which
probably occurred $\sim 2\times 10^8$ yr ago \citep{gn96}. An
examination of the spatial and velocity distribution of future
objects, similar to \hipass\ J1712-64, will ultimately demonstrate
their relationship, if any, with the Magellanic System.

\section{DISCUSSION}
\label{sec:dis}

Although there have been many previous searches for extragalactic \HI\ 
in optically blank fields \citep{fis81a,bri90,bar97,sch98}, there has
been little success, implying that massive \HI\ `protogalaxies' are
rare at the present epoch. However, \hipass\ J1712-64 has a very low
\HI\ mass, $1.7\times 10^7$\ \Msun, implying that similar objects may
exist in abundance and may have escaped previous detection through
lack of sensitivity. Some limits exist from the Cen~A \hipass\ results
of \citet{banks99}, who find 10 new group members, all with
corresponding optical counterparts, in the 600 deg$^2$ they surveyed
around the Cen~A group.  However, their sensitivity limit is $\sim
10^7$\ \Msun\, so it is possible that slightly deeper observations
\citep{dis99} may reveal more such objects. The \hipass\ observations
of the south celestial cap, Decl.  $<-62\arcdeg$ \citep{kws99},
comprise 2400 deg$^2$ of sky, but contains only a few other candidate
`protogalaxies', which are awaiting follow-up CCD data.  However,
\hipass\ J1712-64 is the most significant detection in terms of flux
density and angular size.  Nevertheless, for there to exist large
numbers of similar objects would require them to have \HI\ masses
$<10^7$\ \Msun.

Interestingly, a system with a similar velocity and \HI\ morphology,
ZOA J1616-55, was found in a search for galaxies in the Zone of
Avoidance (ZOA) near $\ell=325\arcdeg$ \citep{sta98}. This 
system has a similar distance and \HI\ column density to \hipass\ 
J1712-64 but a higher \HI\ mass ($9\times 10^7$\ \Msun) and diameter
(86 kpc).  Although lying in the ZOA, the estimated blue absorption is
only $A_B=2.7$ mag, so the lack of an optical or IR counterpart led
\citet{sta98} to conclude it was a galaxy pair of low optical surface
brightness.  J1712-64 is only 11\arcdeg\ (projected separation 600 kpc)
from J1616-55 and 19\arcdeg\ (projected separation 1.1 Mpc) from the
massive Circinus galaxy. Possibly, all three are part of a loose southern
extension to the Cen~A group. In the HVC hypothesis, the alternative
is that Circinus is unrelated and both J1712-64 and J1616-55 have
a Magellanic origin.

From a study of the Ly$\alpha$ forest lines and a VLA \HI\ study,
\citet{sps99} suggest that, at $z=0$, there may exist clouds, or
sheets, of ionized gas $\sim 1$ Mpc in extent which contain $\sim 30$
times more baryons than the neutral gas confined to known galaxies.
Objects such as J1712-64 may be the most easily visible manifestations
of these large primordial clouds. So, although the neutral gas content
in J1712-64 is hardly enough to make a star cluster, let alone a
galaxy, there may be significantly more baryons in the region
available to form a galaxy if they were able to cool. The possibility
that the dynamical mass of the J1712-64 system is $\sim 10$ times
greater than the \HI\ mass suggests there may be dark matter of some
sort associated with the system.

Isolated \HI\ clouds can be used to set stringent limits on the local
metagalactic background radiation. Deep narrow-band CCD images of the
H$\alpha$ line in HI 1225+01 and the Leo ring place limits of $J_0 <
3-8 \times 10^{-23} \, \erg$ for quasar-like light
\citep{don95}. H$\alpha$ observations of J1712-64 may also help
provide strong constraints on this background. Such observations would
also be useful in providing a lower limit to the distance to J1712-64
in light of the observation that many nearby HVCs have faint but
detectable H$\alpha$ emission probably resulting from the Galactic
ionization field \citep{bhm99}.

The mean velocity dispersion of the NE component of J1712-64 appears to be
low, $\sigma = 4$\ \kms\ (Table~\ref{table1}). This gives an upper
limit to the kinetic temperature of $T<2000$ K. The existence of cool
gas would be interesting as this requires that there are densities
high enough to allow cooling and may therefore indicate that
star-formation is about to commence.  In order to search for cool gas,
we examined the ATCA \HI\ spectrum of PKS 1708-648 which is 19\farcm4
from the center of the J1712-64. However, we were unable to
detect any unambiguous evidence for absorption. This is probably not
significant as the continuum source is well beyond the last measured
contour of \hipass\ J1712-64.

\section{SUMMARY}
\label{sec:summ}

An isolated \HI\ cloud, \hipass\ J1712-64, has been found during the
course of the \hipass\ survey. It appears to be a nearby extragalactic
cloud lying beyond the outer fringes of the Local Group at a distance
of $\sim 3.2$ Mpc, though an alternative explanation is that of an HVC 
with unique properties. Optical limits imply a
surface brightness, corrected for obscuration of $\mu(B)>27$ mag
arcsec$^{-2}$ and an \HI\ mass-to-light ratio $M_{HI}/L_B>24
\Msun/L_{\odot}$ for a putative optical counterpart of diameter
1\arcmin. The system appears to have two components, separated by 9
kpc, with a bridge of gas between them.  The mean velocity dispersion of
the NE component is low, $\sim 4$\ \kms\ implying that \hipass\
J1712-64 contains cool gas ($T<2000$ K).  However, it appears to be
dynamically stable against star-formation.

Although the \HI\ mass is low, $\sim 1.7\times 10^7$\ \Msun\ (for a
distance of 3.2 Mpc), the velocity differences within the system
suggest that there may be an extended dark matter halo if the system
is bound.  Is \hipass\ J1712-64 important from a cosmological point of
view? This depends on whether similar objects exist in great numbers,
whether they represent the cool, high-density baryonic peaks of
low-amplitude overdensities, whether such objects are the building
blocks of galaxy and star cluster formation, whether they are simply
the left-overs of these processes, and whether they are indeed
extragalactic.  Future \hipass\ observations will cast some light on
the frequency of their occurrence, and their distribution on the sky.



\acknowledgments

We are grateful to Michael Brown for assistance with analysis of the
CCD data, using SExtractor, to Emma Ryan for assistance with the ATCA
observing, to the staff at the Parkes telescope, and to members of the
ZOA observing team for assistance with observations. P. Knezek
acknowledges partial support by a grant from NASA administered by the
American Astronomical Society.





\clearpage



\plotone{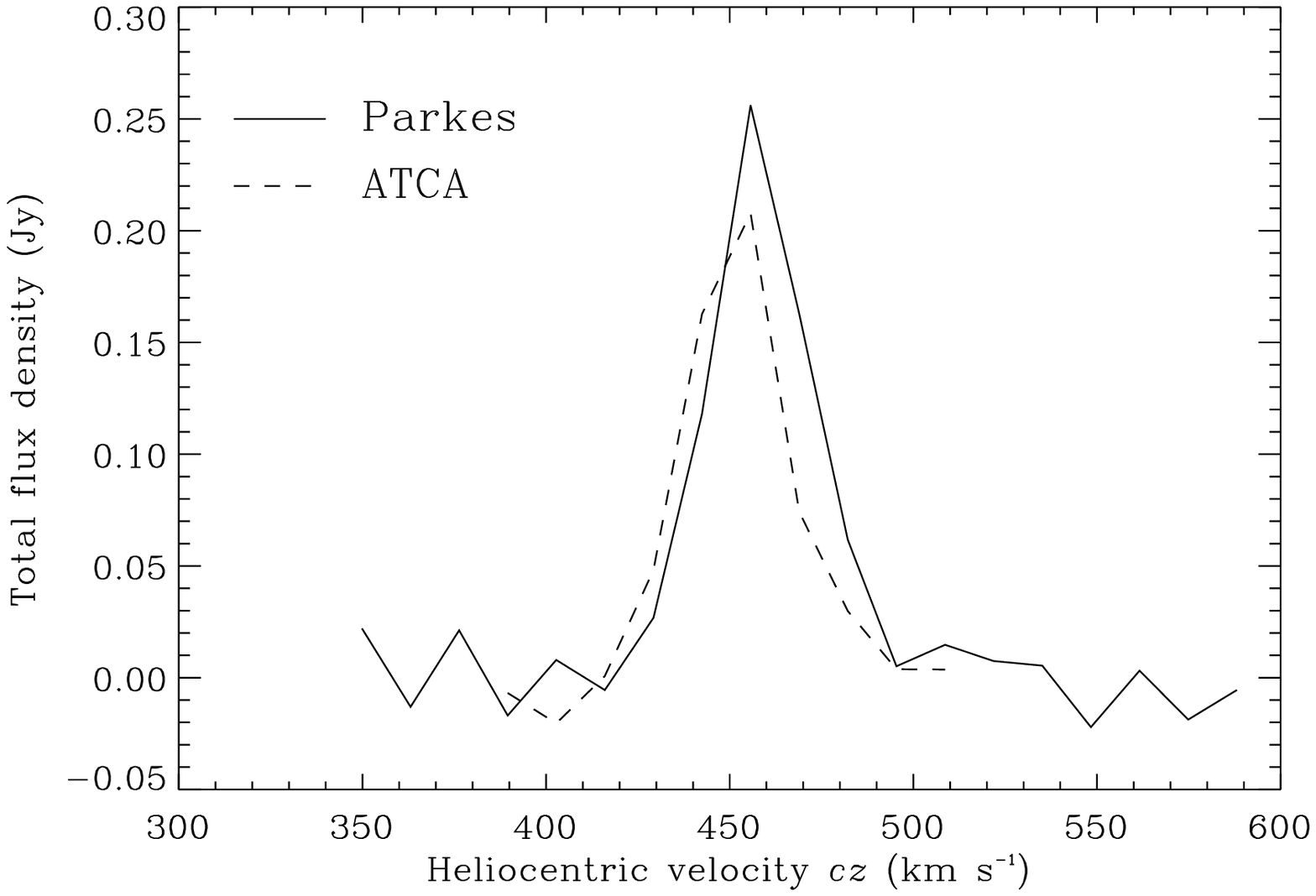}

\figcaption[Kilborn.fig1.ps]{ Spatially integrated \HI\ spectra of \hipass\ 
  J1712-64 from the Parkes/\hipass\ data (solid line) and the ATCA
  (dashed line).  The ATCA appears to recover $\sim 80\%$ of the total
  flux density.  Both spectra are derived by spatially integrating
  over a box centred on the mid-point of J1712-64.  The size of the
  box was $44'$ and $16'$ for the Parkes and the ATCA data
  respectively.  The ATCA data has been smoothed to the spectral
  resolution of the Parkes data. No clipping of the data points was
  used.
\label{f:spec}}

\plotone{Kilborn.fig2.ps}

\figcaption[Kilborn.fig2.ps]{\HI\ channels maps for \hipass\ 
  J1712-64 from the Hanning-smoothed ATCA data. The mean heliocentric
  velocity is given in the top left hand corner of each map. The
  separation of plotted channels is $6.6\ \kms$.  The contours plotted
  are $-5$, 5, 10, 15, 20, 25 and 30\ \mJy\ per beam.
\label{f:chan}}

\plotone{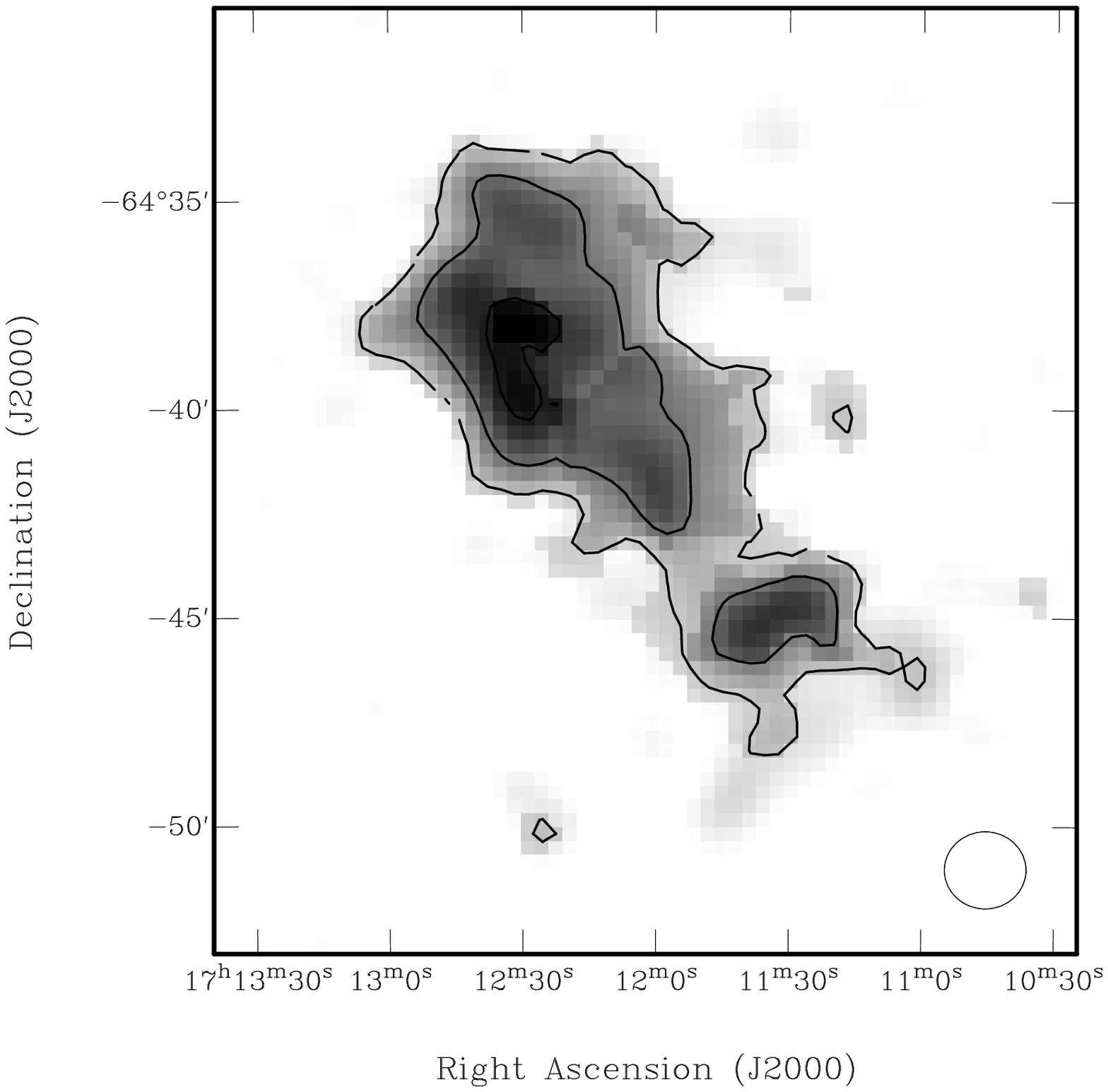}

\figcaption[Kilborn.fig3.ps]{\HI\ column density map of \hipass\ 
  J1712-64 from the combined ATCA/Parkes dataset.  The contours
  represent increments of $1 \times 10^{19}\ \acmsq$ to a maximum
  value of $3.5 \times 10^{19}\ \acmsq$.  The grey scale is a linear
  representation of the column density.  The FWHM of the final beam
  shown in the lower right hand corner is $2\farcm0 \times
  1\farcm9$. \label{f:mom}}

\plotone{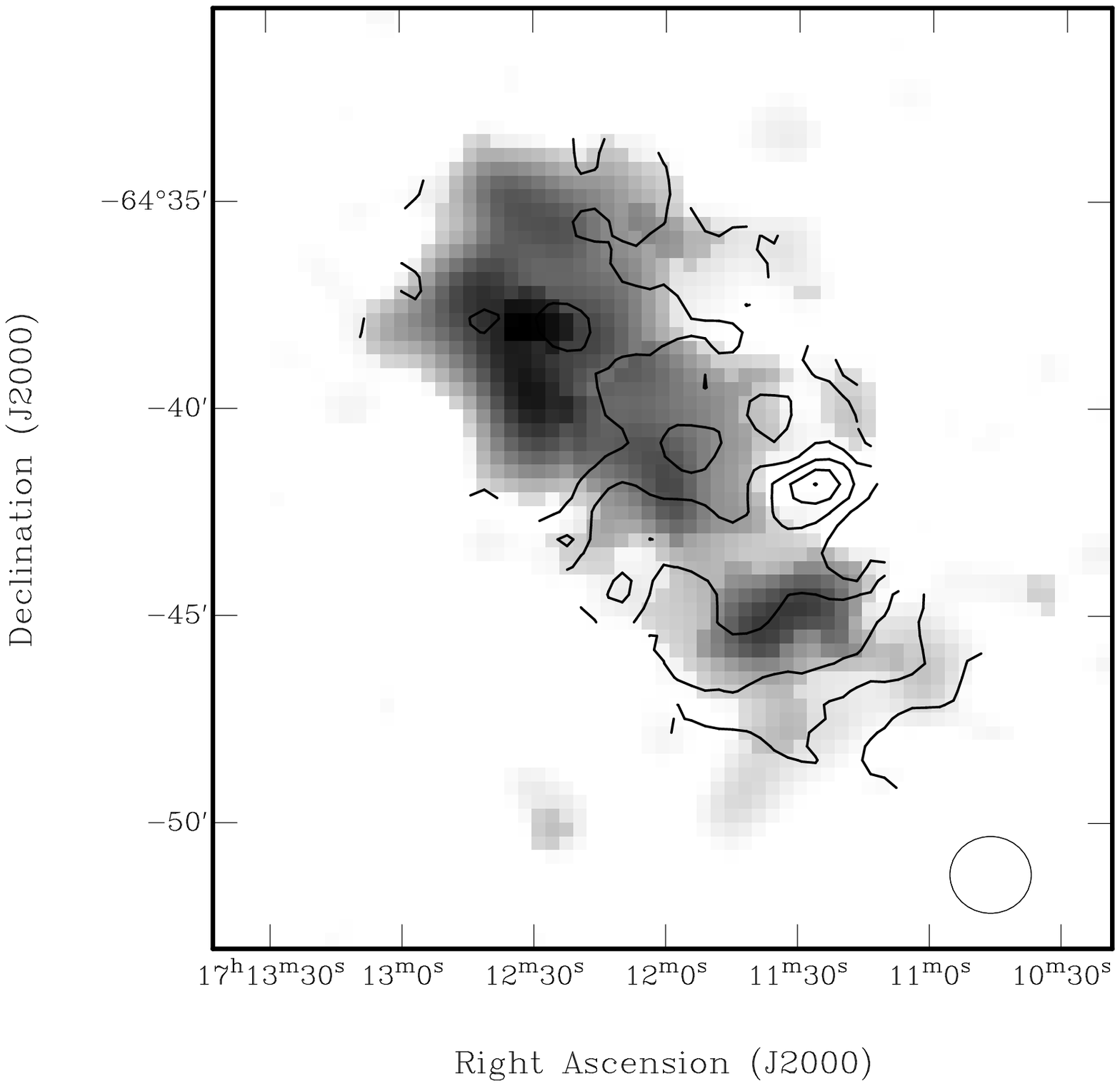}

\figcaption[Kilborn.fig4.ps]{\HI\ velocity field of \hipass\ J1712-64 from
  the ATCA data, superimposed on the column density image (from the
  combined ATCA and Parkes data).  The contours are at intervals of 5\ 
  \kms\ and range from 450\ \kms\ in the northeast to 475\ \kms\ in
  the southwest. The FWHM of the final beam shown in the lower right
  hand corner is $2\farcm0 \times 1\farcm9$. \label{f:mom1}}

\plotone{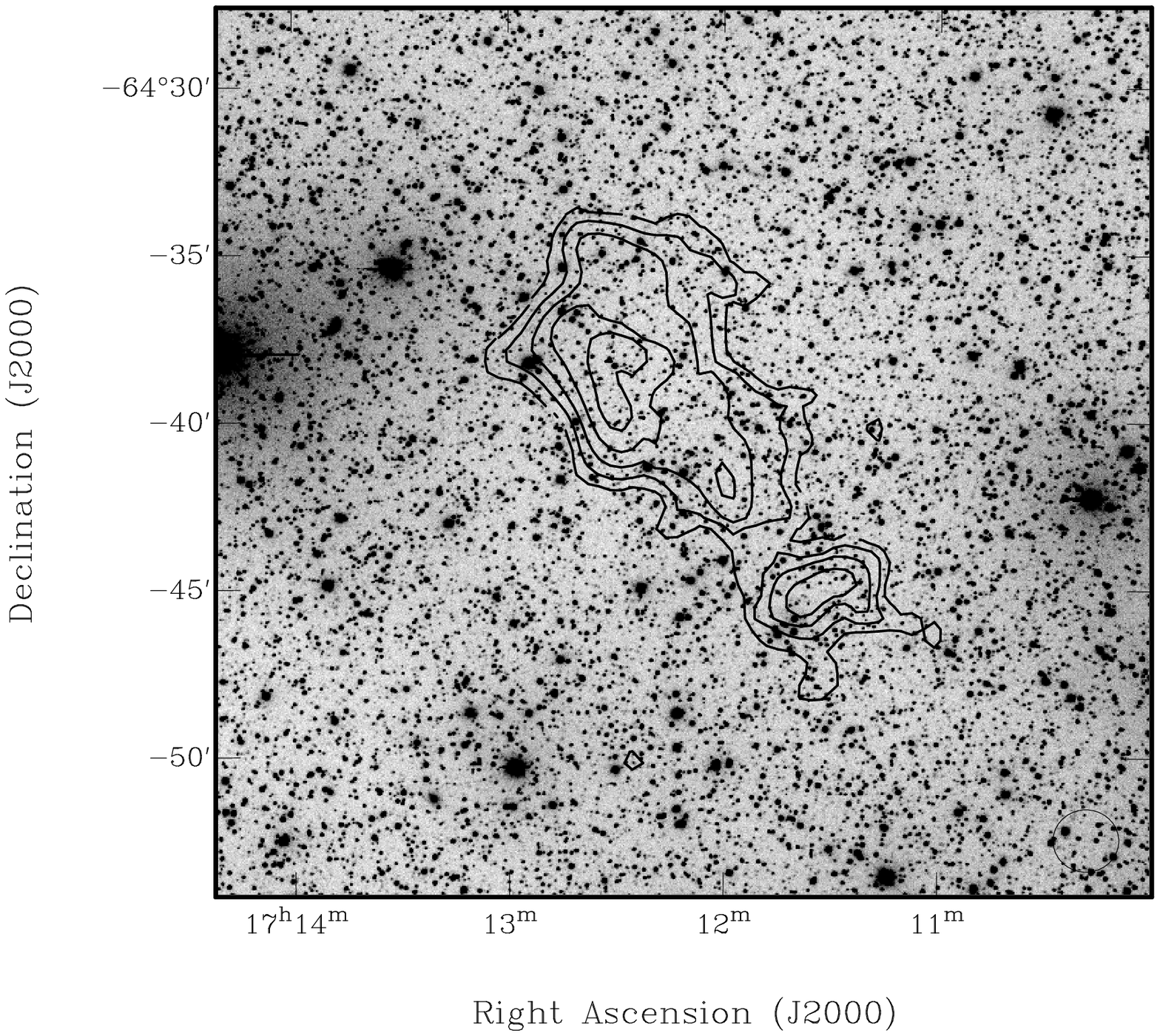}

\figcaption[Kilborn.fig5.ps]{Combined B and R-band CCD optical images
  for \hipass\ J1712-64, with Parkes+ATCA \HI\ contours superimposed.
  \label{f:olay}}

\plotone{Kilborn.fig6.ps} 
\figcaption[Kilborn.fig6.ps]{Number counts of stars in
  the central region of the $B$-band image, and two surrounding
  regions.  The star symbols show star number counts the region where
  we would expect to see an excess of stars. No systematic excess is
  seen.\label{f:hist}}

\plotone{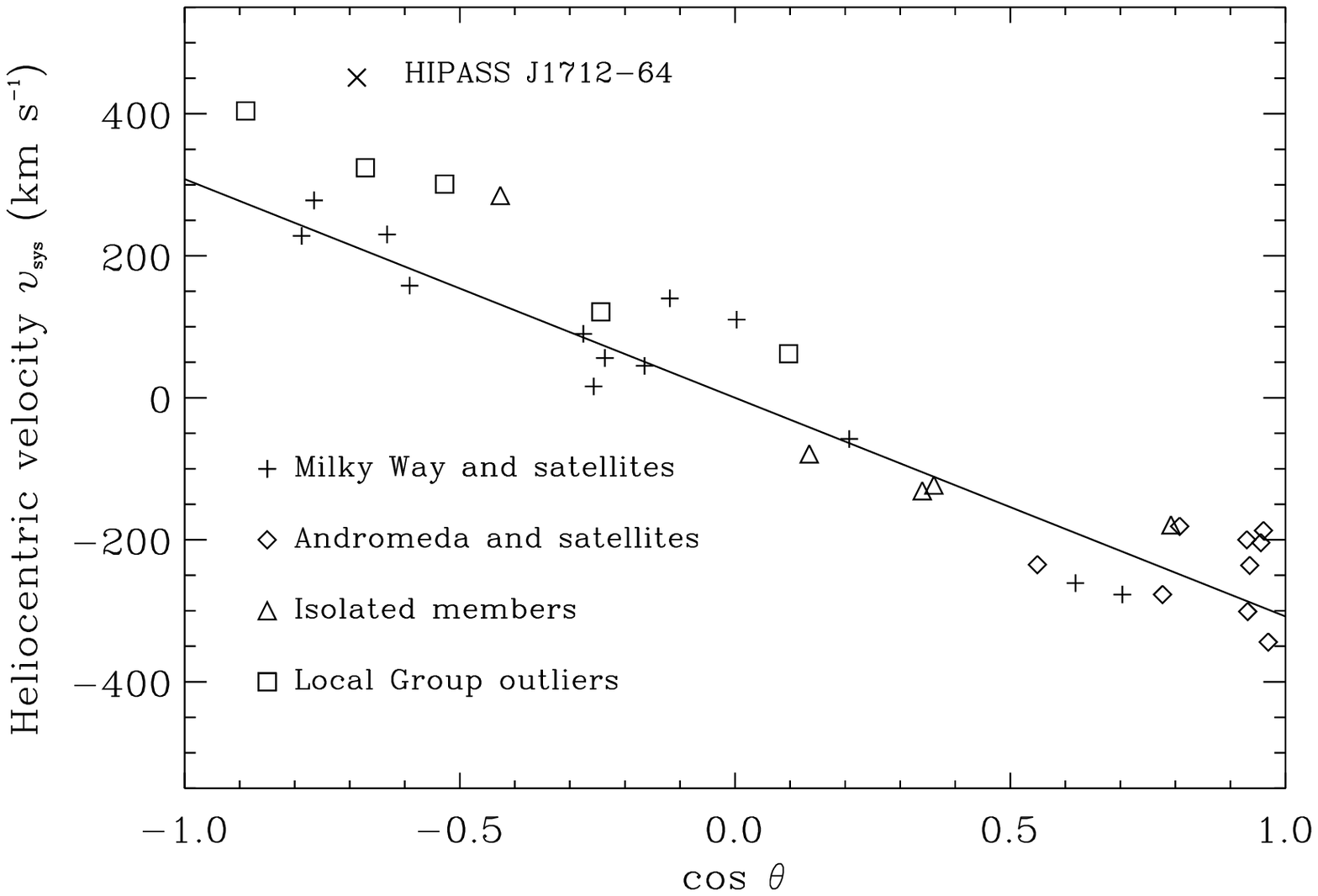}

\figcaption[Kilborn.fig7.ps]{Velocities of Local Group members and outliers
  plotted against the cosine of the angle $\theta$ from the apex of
  solar motion relative to the Local Group barycenter \citep{yts77}.
  \hipass\ J1712-64 has the highest heliocentric and Local Group
  velocity, placing it beyond the Local Group outliers, which are
  themselves only considered to be distant potential members
  \citep{gre97}.\label{f:lg}}





\clearpage

\begin{deluxetable}{lccc}
\footnotesize
\tablecaption{Measured \HI\ Parameters for \hipass\ J1712-64 and its
components\label{table1}}
\tablewidth{0pt}
\tablehead{
\colhead{ } & \colhead{NE Component}   & \colhead{SW Component} & \colhead{J1712-64}  }
\startdata
& \\
{\underline {ATCA + Parkes}} \\
& \\
RA (J2000)              & $17^{\rm h}12^{\rm m}35^{\rm s}$ &
                          $17^{\rm h}11^{\rm m}35^{\rm s}$  \\
Dec (J2000)             & $-64\arcdeg38\arcmin12\arcsec$ &
                          $-64\arcdeg45\arcmin11\arcsec$  \\
$\ell$                  & $326\fdg65$ & $326\fdg48$ \\
$b$                     & $-14\fdg63$ & $-14\fdg61$ \\
Size                    &$11\farcm4 \times 8\farcm2$    &$4\farcm9
\times 4\farcm1$        & $15\farcm9 \times 8\farcm2$\\
PA                      &$32\arcdeg$    &$125\arcdeg$   &$32\arcdeg$\\
Peak column density     &$3.5\times 10^{19}\ \sqcm$       &$2.8\times 10^{19}\ \sqcm$\\
Total flux density      &$4.7\ \Jykms$    &$1.3\  \Jykms$   &$7.0\  \Jykms$\\
& \\
{\underline{ATCA}}\\
& \\
Heliocentric velocity, $v_{\rm sys}$        & $451\  \kms$& $464\  \kms$ \\
50\% velocity width, $W_{50}$             & $11\  \kms$ & $20\  \kms$\\
Velocity dispersion, $\sigma$             & 4 \kms \\

\enddata


\end{deluxetable}

\clearpage

\begin{deluxetable}{lccc}
\footnotesize
\tablecaption{Derived Parameters for \hipass\ J1712-64 and its
components\label{table2}}
\tablewidth{0pt}
\tablehead{
\colhead{ } & \colhead{NE Component}   & \colhead{SW Component} & 
\colhead{J1712-64} }
\startdata

Heliocentric velocity, $v_{\rm sys}$        & $451\ \kms$ & $464\ \kms$ \\
Dynamical LSR velocity, $v_{\rm lsr}$       & $449\ \kms$ & $462\ \kms$ \\
Galactocentric velocity\tablenotemark{a}, $v_{\rm gsr}$ & $332\ \kms$ & 
                                          $344\ \kms$ \\
Local Group velocity\tablenotemark{b}, $v_{\rm LG}$ & $239\ \kms$ & 
                                         $251\ \kms$ \\
Distance\tablenotemark{c}                & 3.2 Mpc & 3.2 Mpc & 3.2 Mpc\\
Spatial size             & $11\ \kpc\times 8\ \kpc$ &$5\ \kpc\times 4\ \kpc$ &
                           $15\ \kpc\times 8\ \kpc$ \\
\HI\ Mass,  $M_{HI}$     & $1.1 \times 10^7\ \Msun$ & $3 \times 10^6\ \Msun$&
                           $1.7 \times 10^7\ \Msun$ \\
Dynamical mass\tablenotemark{d}, $M_{T}$   & $\sim 3 \times 10^7\ \Msun$ &   &
                                   $>1.5 \times 10^8\ \Msun$ \\
Blue Luminosity, $L_B$ &   &  & $<7 \times 10^5 L_{\odot}$ \\
$M_{HI}/L_B$           &   &  & $>24\ \Msun/L_{\odot}$ \\
\enddata


\tablenotetext{a}{$v_{\rm gsr}=v_{\rm lsr}+220 \sin \ell \cos b$.}
\tablenotetext{b}{Using the formula of \citet{yts77}.}
\tablenotetext{c}{Assuming an identical distance for the NE and SW components 
and using $d=v_{\rm LG}/H_{\circ}$ with $H_{\circ}=75$ km~s$^{-1}$~Mpc$^{-1}$.}
\tablenotetext{d}{See text.}

\end{deluxetable}


\clearpage


\end{document}